\documentclass[utf8]{frontiersFPHY} 

\setcitestyle{square} 
\usepackage{url,hyperref,microtype,subcaption}
\usepackage[onehalfspacing]{setspace}

\usepackage{graphics}
\usepackage{amsgen,amsbsy,amsopn,amscd,amsmath,amsfonts,amssymb}

\newcommand{\mb}{\mathbf}

\newcommand{\mc}{\mathcal}

\newcommand{\tx}{\textit}
\newcommand{\lb}{\lambda_\text{deB}}
\newcommand{\lmf}{\lambda_\text{mf}}
\newcommand{\pself}{P_\text{SI}}
\newcommand{\beq}{\begin{equation}}
\newcommand{\eeq}{\end{equation}}
\newcommand{\bdi}{\begin{displaymath}}
\newcommand{\edi}{\end{displaymath}}
\newcommand{\beqn}{\begin{eqnarray}}
\newcommand{\eeqn}{\end{eqnarray}}

\newcommand{\f}{\frac}

\newcommand{\MS}{M_{\odot}}

\newcommand{\vc}{v_{\rm circ}}

\def\keyFont{\fontsize{8}{11}\helveticabold }
\def\firstAuthorLast{T. Rindler-Daller} 
\def\Authors{Tanja Rindler-Daller\,$^{1,2,*}$} 
\def\Address{$^{1}$Institut f\"ur Astrophysik, Universit\"atssternwarte Wien, \\Fakult\"at f\"ur Geowissenschaften, Geographie und Astronomie, Universit\"at Wien,\\ T\"urkenschanzstrasse 17, A-1180 Vienna, Austria\\ 
$^{2}$ Wolfgang Pauli Institut, Oskar-Morgenstern-Platz 1, A-1090 Vienna, Austria}
\def\corrAuthor{Corresponding Author}
\def\corrEmail{tanja.rindler-daller@univie.ac.at}

\begin{document}
\onecolumn
\firstpage{1}

\title[Gross-Pitaevskii theory and DM halos]{On particle scattering in Gross-Pitaevskii theory and implications for dark matter halos} 

\author[\firstAuthorLast ]{\Authors} 
\address{} 
\correspondance{} 

\extraAuth{}

\maketitle

\begin{abstract}

Bose-Einstein-condensed dark matter (BEC-DM), also called scalar field dark matter (SFDM), has become a popular alternative to the standard, collisionless cold dark matter (CDM) model, due to its long-held potential to resolve the small-scale crisis of CDM.
 Halos made of BEC-DM have been modelled using the Gross-Pitaevskii (GP) equation coupled to the Poisson equation; the so-called GPP equations of motion. These equations are based on fundamental microphysical conditions that need to be fulfilled in order for the equations to be valid in the first place, related to the diluteness of the DM gas and the nature of the particle scattering model. We use these conditions in order to derive the implications for the BEC-DM parameters, the 2-particle self-interaction coupling strength $g$ and the particle mass $m$. We compare the derived bounds with the constraint that results from the assumption of virial equilibrium of the central cores of halos, deriving a relationship that connects $g$ and $m$. We find that the GPP conditions are greatly fulfilled, for BEC-DM particle masses of interest, if such models also obey the virial condition that turns out to be the strongest constraint. We also derive the implications for the elastic scattering cross section (per particle mass) in BEC-DM halos, based on the scattering model of GPP, and find a huge range of possible values, depending on the self-interaction regime. We put our results into context to recent literature which predicts sub-kpc core size in BEC-DM halos. 
\tiny
 \keyFont{ \section{Keywords:} cosmology, Bose-Einstein-condensed dark matter, galactic halos, Gross-Pitaevskii theory, particle scattering} 
\end{abstract}

\section{Introduction}\label{intro}

In recent years, alternative models to the cold dark matter (CDM) paradigm have been studied for multiple reasons: i) the hitherto non-detection of CDM candidates such as ``weakly interacting, massive particles (WIMPs)'', or the quantum-chromodynamical (QCD) axion; ii) the small-scale crisis of CDM structure formation, such as ``cusp-core problem'', ``missing-satellite problem'' or ``too-big-to-fail problem'', established by astronomical observations, which could be interpreted to point to a minimum structure formation scale which is larger than that predicted by CDM (for reviews see e.g. \cite{2017ARA&A..55..343B,2019A&ARv..27....2S}); and finally iii) the emergence of a plethora of dark matter (DM) candidates in new theories beyond the standard model of particle physics.

The models that we will consider in this paper belong to a large family of such alternative candidates, so-called Bose-Einstein-condensed dark matter (BEC-DM), introduced around the 2000s by several authors, see e.g. \cite{Sin94,Peebles2000,Goodman2000,sf4,sc19,2000PhRvL..85.1158H,Mielke2006}. Ultralight bosonic DM with particle masses of $m \sim (10^{-23} - 10^{-18})$ eV/$c^2$ belongs to this category, which have been studied under various names: ``scalar field dark matter (SFDM)'', ``fuzzy dark matter (FDM)'', ``wave dark matter'', ``ultralight axion DM''. Reviews on SFDM can be found e.g. in \cite{rev1,UL2019}, on axions e.g. in \cite{rev3}. 
BEC-DM has become popular with astrophysicists, because its dynamics can be distinctive to CDM on galactic scales, possibly overcoming the CDM small-scale crisis. However, this feature depends upon the number of available particle parameters of the model and the respective values they can take. Thus, these parameters are subject to observational constraints, and many have been derived over the years, using different astronomical scales and probes, see e.g. \cite{2015PhRvD..91j3512H,free_const3,sc20,MN19,Nadler21,Safarzadeh_2020,sc22,sc23} for models which have only the particle mass as a free parameter. Recent constraints on models which include a strongly repulsive 2-particle self-interaction (SI), also called ``SFDM-TF'' where ``TF'' stands for ``Thomas-Fermi'' regime, can be found e.g. in \cite{LRS14,LRS17,Fan,Dev,Shapiro21,Hartman21,Foidl}. 

In this paper, we investigate the 2-particle SI in BEC-DM from the quantum scattering perspective that underlies the commonly adopted Gross-Pitaevskii framework, and consider the implications that arise for the BEC-DM particle parameter space.  

For simplicity, we make the usual assumption in that we consider a single species of spin-0 bosons of mass $m$ that forms a Bose-Einstein condensate (BEC), and makes up all of the cosmological DM abundance. We note that mixed-DM models have been also considered, albeit more rarely, where the energy density of DM is shared between BEC-DM and, say, a CDM component, see e.g. \cite{2015PhRvD..91j3512H}. 
The particles of BEC-DM are considered fundamental without internal structure, and we adopt this premise here, as well. In the context of this paper, that means we disregard the possibility of inelastic collisions or bound states between these particles.
Also, we assume that there is no coupling to the particles of the standard model (though there may have been some coupling in the very early Universe, upon the emergence of BEC-DM).

All these assumptions are encoded in the so-called
nonrelativistic Gross-Pitaevskii-Poisson (GPP) system of equations for the complex-valued BEC-DM wavefunction $\psi(\mb{r},t)$, whose application is widely accepted in the community (see references above). The GPP framework is appropriate as a description of galactic halos made of BEC-DM, where both gravitational and
velocity fields are small enough that they can be treated in the Newtonian regime.

The GPP equations of motion are as follows:
\begin{equation} \label{gp}
 i\hbar \frac{\partial \psi(\mb{r},t)}{\partial t} = -\frac{\hbar^2}{2m}\Delta \psi(\mb{r},t) + m\Phi(\mb{r},t) \psi(\mb{r},t) +
 g|\psi(\mb{r},t)|^2\psi(\mb{r},t),
 \eeq
 \beq \label{poisson}
  \Delta \Phi(\mb{r},t) = 4\pi G m |\psi(\mb{r},t)|^2.
   \end{equation}
Since we assume that all of the DM is in BEC-DM,
the mass density of DM particles is given by $\rho(\mb{r},t) = m n(\mb{r},t)$, where the number (probability) density is assigned to the wavefunction squared,
$n(\mb{r},t) = |\psi|^2(\mb{r},t)$, according to the Born rule.  

$\Phi(\mb{r},t)$ is the gravitational
potential of the system under consideration, which couples to the mass density via the Poisson equation (\ref{poisson}). The last term in
(\ref{gp}) describes the low-energy elastic scattering of the DM
particles in this model, parametrized by the coupling constant
 \beq \label{coupling}
  g = 4\pi \hbar^2 \f{a_s}{m},
   \eeq
with the 2-particle s-wave scattering length $a_s$, whose sign
determines whether the self-interaction (SI) is repulsive, $g > 0$, or attractive, $g < 0$. If SI is neglected, $g=0$, the above system of equations is also known as Schr\"odinger-Poisson (SP) system\footnote{The GP equation in (\ref{gp}) is formally a nonlinear Schr\"odinger equation, given the nonlinear SI-term. In addition, nonlinearity comes in via the coupling of GP to the Poisson equation (\ref{poisson}), to describe self-gravitating systems. While the GP equation resembles formally the linear Schr\"odinger equation if $g=0$, the overall Schr\"odinger-Poisson system is nonlinear in any case. Moreover, the wavefunction $\psi$ of GP is interpreted as the ``c-number'' of a highly occupied many-body quantum state, in the limit of large particle number, thus it is not the wavefunction of a single particle, but rather that of the macroscopic condensate itself. The GP model is effectively a mean-field description.}.
The significance of the above scattering model, its implications for BEC-DM parameters and comparison with other constraints will
be the subject of this paper.

By applying the GPP framework in (\ref{gp}-\ref{poisson}), we appreciate that the wave nature of the DM bosons cannot be neglected on astronomical or galactic scales anymore, i.e. their quantum properties can affect observational phenomenology, depending upon the parameters of the model. The GPP framework can be used for any DM bosons which form a BEC, whether they be ultralight $m \sim 10^{-22}$ eV/c$^2$ appropriate for FDM, or whether they be QCD axions with $m \sim (10^{-6}-10^{-5})$ eV$/c^2$.  
However, the GP equation (\ref{gp}) has been applied to BEC-DM in previous literature without consideration of the actual assumptions that underlie the equation in the first place. We will close this gap by determining the DM parameter space of $m$ and $g$ for which we can apply the GP description, the implications for the scattering cross section, and how these fit into dynamical constraints.

Some comments are in order before we proceed. First, different
cosmological production mechanisms for BEC-DM in the very early
Universe have been envisaged.
However, the
question as to whether the effective low-temperature and
nonrelativistic description of equ.(\ref{gp}-\ref{poisson})
represents a proper mathematical limit for BEC-DM, which originates via
different cosmological scenarios, has not been studied in detail and is also beyond the scope of this paper. Here, we merely start with the premise that GPP can be used as a phenomenological framework to study BEC-DM on galactic halo scales, along the lines of previous investigations in the literature.
Second, many popular particle models for BEC-DM involve ultralight axions (ULAs), which are usually pseudo-scalar fundamental particles described via real fields. As such, ULAs can self-annihilate. However, the rate of this self-annihilation is very low; hence it is usually neglected and a complex wavefunction $\psi$ is adopted, as well. (If the particle number is conserved -either exactly or approximately- the system (\ref{gp}-\ref{poisson}) obeys a normalization condition, such that e.g. $N = \int |\psi|^2 d^3\mb{r}$ for the wavefunction $\psi$ of a given galactic halo with total number of bosons $N$.)
More generally, we stress that specific particle models of BEC-DM may impose further relationships between the underlying particle parameters. In the above GPP framework, $m$ and $g$ are independent parameters (see more below), while e.g. ULAs obey a relationship between mass $m$ and their so-called decay constant $f$. The latter can be related to a coupling parameter akin to $g$. So, the GPP ``mean-field'' framework is meant to encompass a broad family of models, whose members behave very closely dynamically, and yet differ with respect to the individual particle model. To determine whether the differences in the latter play a role on astronomical scales, or not, requires a case-by-case analysis which is not subject of this work. We will focus on the generic SI-term in equation (\ref{gp}), without consideration of how it arises from given particle Lagrangians\footnote{There are two prominent possibilities why a $|\psi|^4$-term in the Lagrangian, that gives rise to the $|\psi|^2\psi$-term in the equation of motion, arises. On the one hand, $\psi$ could serve as a Landau ``order parameter'' that describes the BEC state after a phase transition. In that case, it obeys translational symmetry and in the expansion of the (free) energy (functional) only even powers of $\psi$ occur, with the usual truncation after the quartic term. On the other hand, if $\psi$ serves as some relict scalar field that acquires its mass through some vacuum misalignment mechanism, like the QCD axion, the resulting Lagrangian has a cosine potential, which can be expanded and is also usually truncated after the quartic term. As another comment, we add that a quadratic potential can always be absorbed in the definition of the chemical potential of the system. }.   
Third, in the Newtonian regime, the self-consistent coupling of gravity to the mass density
is established via the Poisson equation in (\ref{poisson}). Yet, the smooth global gravitational potential $\Phi$ in the gravitational term $m\Phi \psi$ of the GP equation (\ref{gp}) acts nevertheless akin to an ``external potential'', similarly to magneto-optical trap potentials in typcial laboratory atomic BECs. In other words, $\Phi$ does not ``feel'' the particle SI, described via the SI-term $g |\psi|^2\psi$, because $\Phi$ is basically uniform over the interaction range of the
2-particle SI-potential which is essentially contact-like. In fact, the equation for the SI coupling strength in (\ref{coupling}) arises as a result of assuming a $\delta$-distribution for the SI-potential; see textbook presentations e.g. \cite{pethick_smith_2008,PS}. This situation is very different from the case of, say, a laboratory molecular BEC with dipolar forces, where the 2-boson SI itself is a long-range $1/r$-interaction. 
Here, we exclude such a case and assume throughout that the bosons of BEC-DM have short-range SI.
Finally, it will become clear in due course why we will mostly focus on the cores of galactic halos. It is these (smaller) halo \tx{cores} which serve as minimum-size equilibrium objects in BEC-DM.

This paper is organized as follows. In Section \ref{sec:2}, we derive the microphysical conditions on the bosons of BEC-DM which are implied by the GPP framework. Section \ref{sec:3} includes a derivation of the relation that connects the BEC-DM parameters $m$ and $g$ from the condition of virial equilibrium of halo cores, and comparison with the results of Sec.\ref{sec:2}. In Section \ref{sec:4}, we discuss the implications for the 2-particle elastic scattering cross sections in different collision regimes, including the question of velocity-dependence. Finally, Section \ref{sec:5} contains our conclusions and discussion.

\section{Dark matter and Gross-Pitaevskii theory}
\label{sec:2}

Before we investigate the microphysical constraints on DM particle parameters, which are based upon the GP equation, we need to introduce an important characteristic length scale, namely the de Broglie wavelength   
\beq \label{dB}
    \lb = h/(m v).
 \eeq
Here, $m$ is the mass of the DM boson and $v$ is their ``collective'' velocity, which depends on the environment. If the bosons make up a galactic DM halo in gravitational equilibrium, $v$ is close to its virial velocity (which can be associated with a ``virial temperature'').   
If the bosons are part of uncollapsed ``background matter'', the velocity will be correspondigly lower\footnote{For comparison, in the formation of BECs in the laboratory it is customary to speak of the \textit{thermal} de Broglie wavelength, $\lb^{th} \propto 1/\sqrt{T}$, because the thermodynamic temperature $T$ is associated with the velocity of particles of mean kinetic energy, see \cite{TRDReview} for a discussion.}. As also discussed below, it is the velocity dispersion pressure of random wave motion which gives rise to the collective velocity $v$ in galactic halos (see \cite{Taha21,HuiReview}). As such, this velocity should be used in equ.(\ref{dB}).
However, we will make the simplifying assumption that $v$ in (\ref{dB}) as a proxy for the virial velocity can be approximated by the circular velocity, $\vc = (G M/R)^{1/2}$, with the mass $M$ and radius $R$ of the virialized object in question (typically a halo core in what follows).

A system is 
Bose-Einstein-condensed, if the matter waves of individual particles of size $\lb$ overlap, forming a giant matter wave, within which particles have become indistinguishable.
This picture finds a more precise notion in the condition that the phase-space density be a number of order one (i.e. high), 
\beq \label{phasespace}
n \lb^3 \gtrsim 1,
\eeq
where $n$ is again the number density of particles.
This is a fundamental condition, if DM bosons are to be described by a macroscopic wavefunction $\psi$ that obeys the GPP equations.  
Rewriting (\ref{phasespace}) in terms of fiducial galactic units of interest to us, we have the following condition on the boson mass
\beq \label{massboundphasespace}
m \lesssim 85 ~\text{eV}/c^2 \left(\f{\bar \rho}{\text{GeV}/c^2\text{cm}^3}\right)^{1/4}\left(\f{100 ~\text{km/sec}}{\vc}\right)^{3/4},
\eeq
where the bar indicates a mean mass density, obtained from $n$.
Let us consider two illustrative cases. If we assume 
homogeneous halo cores of radius $\sim (0.1-1)$ kpc, which inclose a (core) mass of, say, $10^8~\MS$, we get $m \lesssim (200-700)$ eV/c$^2$.   
On the other hand, for Milky Way parameters at the solar circle, using $\vc = 220$ km/sec and $\bar{\rho} = 0.3$ GeV/($c^2$cm$^3$) for the mean local DM density\footnote{This value stems from \cite{Bovy12}. Newer studies using GAIA data find very similar estimates of 
$\bar{\rho} \simeq (0.3-0.4)$ GeV/($c^2$cm$^3$) for the local DM density, see \cite{Eilers_2019,deSalas_2019}.}, we have $m \lesssim 35$ eV/$c^2$. This is why Ref.\cite{HuiReview} talks about the regime of ``wave DM'' below this mass scale, as opposed to the particle regime above it (see also \cite{Fan}). Since the phase-space condition shall be fulfilled throughout a given galactic halo (because DM is in a BEC everywhere!), it is thus better to apply the condition in low-density environments, resulting in tighter upper bounds on $m$.

Now, we move to the specific microscopic assumptions that underlie the GP equation, hence are binding for BEC-DM.
It is well known from the theory of interacting quantum
gases\footnote{A short and clear presentation can be found for
instance in \cite{stoof}.} that \tx{two essential conditions} are placed in the
derivation of the GP equation, which can be expressed in terms of
the number density of the gas $n = |\psi|^2$, the range of the
particle interaction\footnote{We follow the cited papers for the notation of the particle interaction range ``$r_0$''. It shall not be confused with a radial quantity related to the halo, such as a core radius. We avoid this confusion by designating the subscript ``$c$'' to halo core properties in the forthcoming.} $r_0$, and the de Broglie wavelength of
the bosons $\lb$, as follows.

The first condition on BEC-DM as a quantum gas is the smallness of the ``gas
parameter'' $nr_0^3$. In physical terms,
 \beq \label{cond1}
 n r_0^3 \ll 1
  \eeq
 describes the
fact that a dilute system is considered. It is assumed that the interaction
range of the 2-particle SI potential is much smaller than
the mean interparticle distance. (In fact, the contact-SI alluded to in Section \ref{intro} has this property.) This assumption also guarantees that the probability of
simultaneous $\mc{N}$-particle interactions with $\mc{N} \geq 3$ is so
low that we need only consider 2-particle processes\footnote{In fact,
the GP description for the wavefunction of the condensate can be
derived from the more general Heisenberg equation for the field
operator by truncating terms higher than first order in this gas
parameter; see e.g. \cite{HT} and references therein.}. Moreover, it allows us to introduce a meaningful notion of mean-free path, which we will need in Section \ref{sec:4}.

The second condition requires that
 \beq \label{cond2}
 r_0/\lb \ll 1
  \eeq
   holds. As in \cite{stoof}, we call $r_0/\lb$ the ``quantum parameter''. 
   Then,
 the contribution due to the relative angular momentum of two interacting particles in the
scattering process can be neglected.
Consequently, it can be shown
 that for low energy \tx{and} for a 2-particle potential $W(\mb{r})$ decreasing
 faster than $r^{-3}$ at infinity, the scattering amplitude $f(k, \mb{n}, \mb{n'})$, with $k$ the wavenumber and
 $\mb{n}, \mb{n'}$ the direction of incoming and scattered particle, respectively, becomes
 isotropic. Furthermore, if the limit $k \to 0$ (or $E \to 0$, respectively) exists, it
 thereby defines the s-wave scattering length according to $\lim_{k\to 0} f(k,\mb{n},\mb{n}') = -a_s$.
   The elastic scattering cross section of
indistinguishable bosons in the limit $k \to 0$ is then energy-independent and given\footnote{As opposed to the result $\sigma_s = 4\pi a_s^2$ for distinguishable particles.} by
 \beq \label{sigma}
  \sigma_s = 8\pi a_s^2.
  \eeq
 Expanding the scattering length in terms of the potential, and
 truncating after the first term (the so-called ``first Born
 approximation''), results\footnote{In general, the Born approximation assumes a sufficiently weak scattering process, where the scattered wave is a small perturbation on the incident plane wave. Nevertheless, one may truncate the potential at higher-order terms, as we will discuss in Sec.\ref{sec:4}. Furthermore, we stress that the scattering process referred to here is a quantum-mechanical treatment in the nonrelativistic limit, where one considers scattering of a particle in a potential, justified as long as the relative velocities are much smaller than the speed of light. This assumption is applied in many calculations of scattering cross sections in self-interacting dark matter in general, see e.g. \cite{SIDM_scatt}. Going beyond this treatment requires a fully-fledged quantum-field theoretical calculation. } in the following expression
  \beq \label{scattlength2}
   a_s = \f{\mu}{2\pi \hbar^2}\int W(\mb{r})d^3r,
    \eeq
     where $\mu = m_1m_2/(m_1+m_2)$ denotes the reduced particle mass
 (for more details on low-energy scattering, see textbook presentations e.g. \cite{joachain,LL,PS}). Thus, $a_s$ is proportional to $m$
 for particles of the same
species, $m_1=m_2=m$. As a result, the
coupling strength $g$ in (\ref{coupling}) does not depend on the particle
mass $m$ and is an \textit{independent} parameter of BEC-DM
within the first Born approximation, apart from $m$. Therefore, with all these conditions in place, the
particle interactions in a fully condensed (or almost completely condensed) BEC-DM
quantum gas can be accurately described by considering only all the
possible 2-particle \tx{s-wave} scattering processes. This allows us to
identify $r_0$ in (\ref{cond1}) and (\ref{cond2}) in the
forthcoming with the scattering length $|a_s|$ (where the sign of $a_s$ can be
positive or negative), since $|a_s| \lesssim r_0$ for
the Born approximation to be valid.

\begin{figure*} 
\begin{minipage}{0.5\linewidth}
     \centering
     \includegraphics[width=8cm]{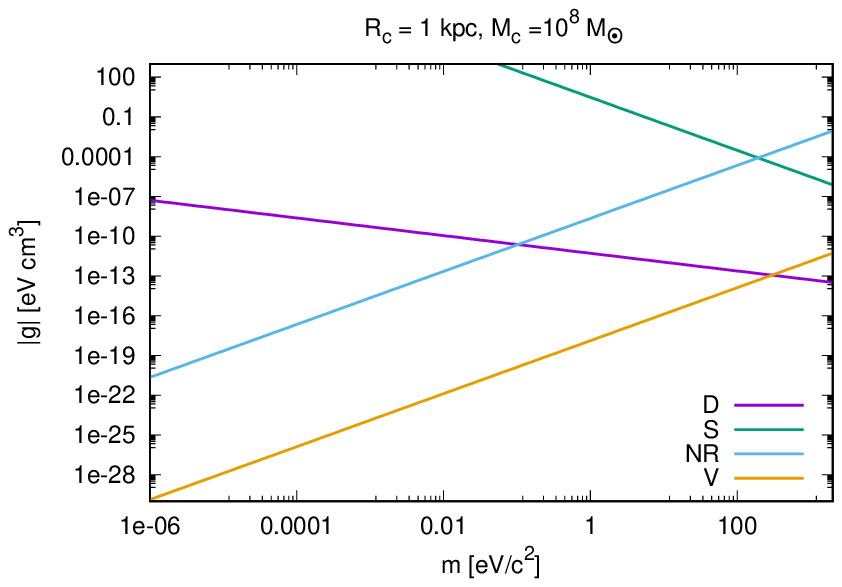}
     \vspace{0.05cm}
    \end{minipage}
    \begin{minipage}{0.5\linewidth}
      \centering
      \includegraphics[width=8cm]{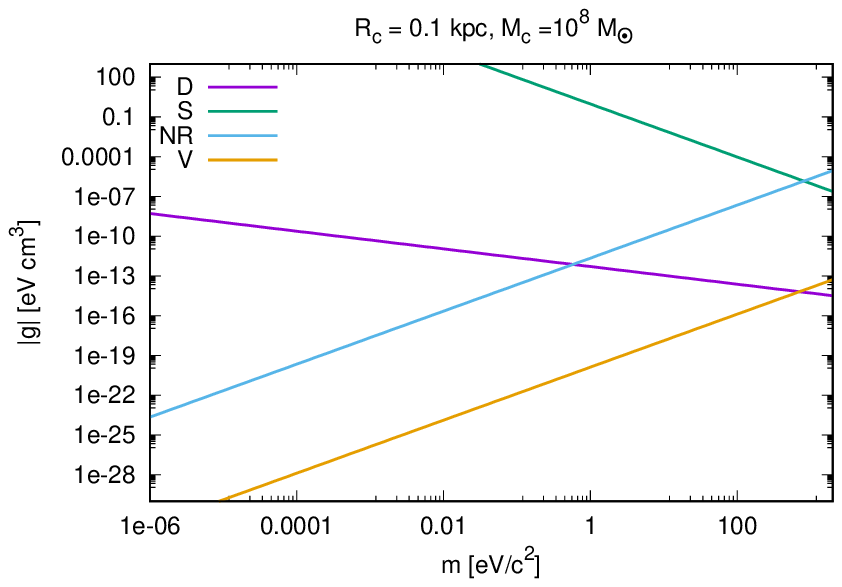}
     \hspace{0.05cm}
    \end{minipage}
 \begin{minipage}{0.5\linewidth}
     \centering
     \includegraphics[width=8cm]{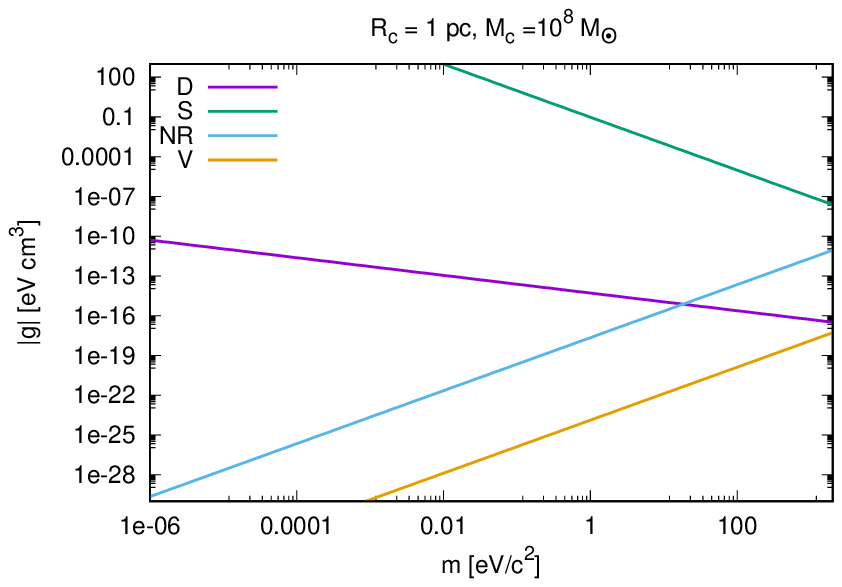}
     \vspace{0.05cm}
    \end{minipage}%
    \begin{minipage}{0.5\linewidth}
      \centering
      \includegraphics[width=8cm]{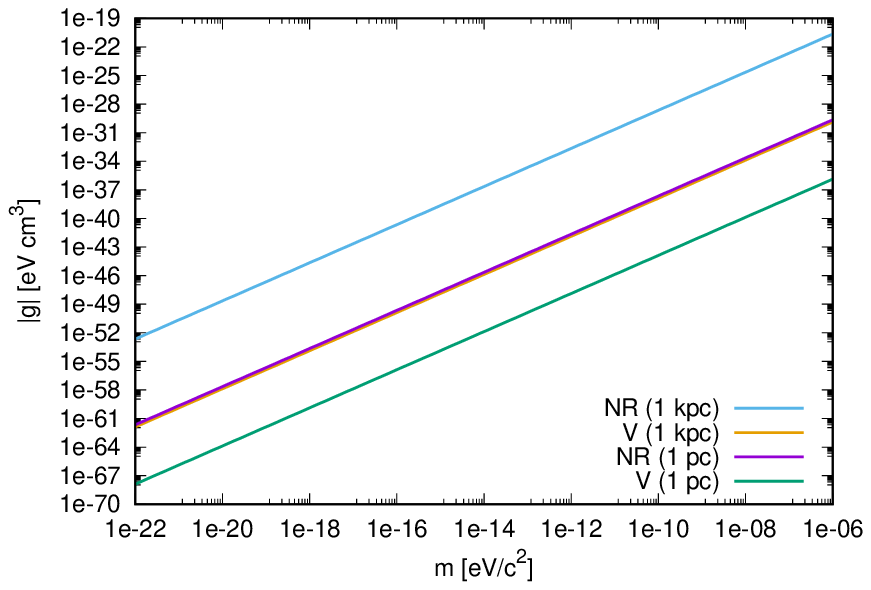}
     \hspace{0.05cm}
    \end{minipage}
 \caption{BEC-DM particle parameter space $(m,g)$ for halos with core mass $M_c = 10^{8}M_{\odot}$, but different core radius: $R_c = 1$ kpc (upper-left panel), $R_c = 0.1$ kpc (upper-right panel) and $R_c = 1$ pc (lower-left panel). Plotted are the respective strong upper bounds on the SI coupling strength $|g|$ (in eV cm$^3$), for a range of particle masses $m \geq 10^{-6}$ eV/$c^2$, from the conditions of
 diluteness 'D' (equ.\ref{gcond1}), s-wave scattering 'S' (equ.\ref{gcond2}) and nonrelativistic pressure 'NR'
 (equ.\ref{pressureless}). The virial relation 'V' (equ.(\ref{tfvirial})) is derived in Section \ref{sec:3}. In each panel, the parameter space below the lowest-lying line fulfills all the bounds shown in turn. A change in $R_c$ shifts the bounds, albeit not strongly. \tx{Lower-right panel:} BEC-DM particle parameter space for a range of boson mass that includes the ultralight regime. As the tightest bounds, only the 'NR' and 'V' lines are shown. We plot them for two cases of core radii, $R_c = 1$ kpc and $R_c = 1$ pc, respectively. By coincidence, the curves for 'V' ($1$ kpc) and 'NR' ($1$ pc) lie visually almost on top of each other. }
 \label{fig1}
\end{figure*}

Of course, in contrast to laboratory examples, the values for the parameters $m$ and $g$ for BEC-DM are unknown.
The boson mass $m$ is a free parameter in many particle models of BEC-DM (such as ULAs), with the common implication that its value is undetermined over many orders of magnitude. The same is mostly true for the 
SI coupling parameter $g$ (or related couplings), although SI is often even neglected from the outset for mere simplicity.
Therefore, experimental and observational constraints need to be placed on both, $m$ and $g$, in order to scan the (dis-)allowed BEC-DM parameter spaces exhaustively. 
Here we ask what the implications are for these parameters from the fundamental conditions just described.
First, it will be useful
to write $a_s$ and $\lb$ in fiducial
units.
Using (\ref{coupling}) and (\ref{dB}), we can
write
  \beq \label{scattlength}
  |a_s| = 2.04\times
  10^{8}\left(\f{m}{\rm{eV}/c^2}\right)\f{|g|}{\rm{eV} \rm{cm}^2},
   \eeq
   where $|g|$ has units of eV cm$^3$,
   and
 \beq  \label{dBfidu}
   \lb = 0.37 ~\rm{cm} \left(\f{m}{\rm{eV}/c^2}\right)^{-1}\left(\f{\vc}{100~\rm{km/sec}}\right)^{-1},
   \eeq
   where $\vc$ stands again for the circular velocity as a characteristic velocity.
Now, the diluteness condition (\ref{cond1}), expressed as $|a_s|^3 n \ll 1$, demands that
 \bdi
 m |g| \ll 4\pi \hbar^2 n^{-1/3},
 \edi
 which we write as a condition on the
 SI coupling strength, for given $m$ and mean mass density $\bar \rho$, according to
 \beq \label{gcond1}
 |g| \ll 5\times 10^{-12} ~ \rm{eV}\rm{cm}^3 \left(\f{m}{\rm{eV}/c^2}\right)^{-2/3}
 \left(\f{\bar \rho}{\rm{GeV}/c^2\rm{cm}^3}\right)^{-1/3}.
  \eeq
The second condition, which is the smallness of the quantum parameter (\ref{cond2}), expressed as $|a_s|/\lb \ll 1$, results in
 \bdi
 m^2 |g| \ll 8\pi^2 \hbar^3/v,
 \edi
 or, for given $m$ and $\vc$, a condition on the coupling strength, as follows
  \beq \label{gcond2}
  |g| \ll 0.63~ \rm{eV}\rm{cm}^3 \left(\f{m}{\rm{eV}/c^2}\right)^{-2}\left(\f{\vc}{100 ~\rm{km/sec}}\right)^{-1}.
   \eeq
 We stress again that these conditions on $|g|$ need not be fulfilled in the early Universe when
 the density or kinetic energy (resp. velocity) of BEC-DM or its precursor was very high, such that the GPP formalism breaks down. However, the above conditions are applicable to galactic
 halos for which each $\bar \rho$ and $\vc$ are nonrelativistic. 
While we noted above that the phase-space condition is better probed in low-density environments to get a more stringent upper bound on $m$, we can use a homogeneous-sphere approximation for the halo cores\footnote{The homogeneous sphere is not too bad an approximation for the central constant DM density cores of dwarf spheroidal galaxies, with numbers as high as 
$\bar{\rho} \lesssim (0.1-0.2)~ M_{\odot}/\rm{pc}^3 \simeq (5-7)$ GeV/($c^2$cm$^3$), see e.g. \cite{Burkert_2015}. However, it fares quite poorly in other environments, such as the local DM density in the Milky Way which would yield a value a factor $\sim 44$ higher than what is estimated, see footnote 2.}, without lack of generality and given the involved orders of magnitude, in order to relate $\bar \rho$ and $\vc$ in a simple manner with fiducial core radii $R_c$ and masses $M_c$.  
Thus, we use the circular velocity at the edge of the halo core, written in fiducial units,
\beq \label{vcirc}
\vc = \sqrt{\f{GM_c}{R_c}} = 20.75 ~\rm{km}/\rm{sec} \left(\f{M_c}{10^8~M_{\odot}} \right)^{1/2} \left(\f{\rm{kpc}}{R_c} \right)^{1/2},
\eeq
and likewise for the assumed constant core density,
\beq \label{rhohom}
\bar \rho = \f{3M_c}{4\pi R_c^3} = 0.90 ~\rm{GeV}/c^2 \rm{cm}^3 \left(\f{M_c}{10^8~M_{\odot}} \right) \left(\f{\rm{kpc}}{R_c} \right)^{3}.
\eeq 
   In Fig.\ref{fig1}, we plot the upper bounds on $g$ for different halo core radii $R_c$ but a fixed choice of core mass of $M_c = 10^{8}~M_{\odot}$, where the right-hand side of (\ref{gcond1}) is labeled as 'D' (``diluteness''), while the right-hand side of (\ref{gcond2}) is labeled as 'S' (``scattering'').  
   Before we comment on the results, we introduce a further condition here, as follows. 
The built-in requirement in the GPP formalism that the bosons be nonrelativistic also implies that any pressure contributions  $P_i$ of BEC-DM must be much smaller than its energy density, $(\sum_i P_i)/\bar \rho c^2 \ll 1$. Hence, these pressures are subject to observational contraints\footnote{Sufficiently small pressures also guarantee the similarity to CDM on scales demanded by cosmological observables, e.g.  the expansion history during structure formation, or the subsequent development of the latter. Still, differences between BEC-DM and CDM have been identified concerning the very early Universe, or the amount of small-scale structure. We do not discuss these here, but see e.g. \cite{LRS14,SY09,TRDReview}.},
as pointed out early on in \cite{Peebles2000} and \cite{Goodman2000}.
The nature of the pressures involved can be more easily seen, once the GP equation has been transformed to an equivalent quantum fluid formulation, revealing a set of continuity and momentum conservation equations \cite{Taha21}. 
The upshot is that the corresponding momentum equation contains the extra pressure terms that affect the dynamics of BEC-DM in ways that can be quite different from CDM, depending on scale. The pressure due to the quantum kinetic energy, the so-called ``quantum pressure'', is related to the physics on scales of $\lb$, and it gives rise to an effective large-scale ``velocity dispersion pressure'' within the envelopes of galactic halos, see \cite{Taha21}. Here, random wave motions take the role of orbital motions of collisionless particles in CDM (which also provides a velocity dispersion pressure in the latter\footnote{In fact, in analogy to the derivation of Jeans equations from a collisionless Boltzmann equation, one can show that a fluid picture is also possible for collisionless CDM or stellar systems, respectively, see \cite{AS,Hensler}. The corresponding Euler momentum fluid equation in that case has a pressure term due to the velocity dispersion of the underlying particles (CDM or stars, respectively).}). The associated energy is very much nonrelativistic, because $\vc \ll c$ for typical dynamical velocities $\vc$ within halos, i.e. the quantum pressure of BEC-DM is much smaller than its energy density.

Moreover, for BEC-DM models with SI-term in the Lagrangian, as is the focus of this paper, another pressure arises. If SI is weak, this pressure may be simply negligible compared to the quantum pressure. If there is a strongly attractive SI, instead, the corresponding pressure is negative and acts in favour of gravity, and the system is prone to gravitational collapse (see footnote 14). 
However, if the pressure is due to a strongly repulsive SI, it acts against gravity and provides a stabilizing effect.
For this pressure, the equation of state has the form of an $(n=1)$-polytrope, $P_\text{SI} = K \rho^{1+1/n}$, where $K=g/2m^2$ is a fixed constant for given BEC-DM model parameters $m$ and $g$ \cite{Peebles2000,Goodman2000}. The requirement on this pressure that $\pself/\bar{\rho} c^2 \ll 1$, then translates into a necessary condition on the coupling strength, for given $m$ and $\bar \rho$,
 \beq \label{pressureless}
  g \ll 2\times
  10^{-9}~ \rm{eV}\rm{cm}^3 \left(\f{\bar \rho}{\rm{GeV}/c^2\rm{cm}^3}\right)^{-1}\left(\f{m}{\rm{eV}/c^2}\right)^2.
 \eeq
Of course, any other weaker SI-pressure must obey this bound, as well.
In Fig.\ref{fig1}, the label 'NR' refers to the right-hand side of (\ref{pressureless}).
Finally, the plots in Fig.\ref{fig1} also include the bound --denoted 'V' -- that follows from the assumption of virial equilibrium of halo cores. Its derivation will be the subject of the next section. 
Now, let us comment on the results of Fig.\ref{fig1}. The upper two panels, as well as the lower-left panel, display a range of boson mass at the high end, chosen as $(10^{-6} - 2000)$ eV/c$^2$; the panels include all bounds mentioned and show results for three different values of the core radius $R_c$, for the same core mass, as indicated in the panel's title. We immediately see that the GPP microphysical conditions can be easily fulfilled, even for this higher boson mass range, because the bounds on $|g|$ are generous. Indeed, we can see that the 'NR' condition, as well as the virial condition 'V', provide tighter constraints in general. Only around $m \sim (0.1-10)$ eV$/c^2$, the 'NR' and 'D' bounds switch their roles, and this barely depends on the core size: above a boson mass of roughly that scale, the diluteness condition 'D' becomes the most demanding
 one of all micro-conditions, i.e. particles with higher $m$ must have a corresponding $|g|$, which is quite smaller than what would be allowed by the 'NR' condition by itself.
 The condition from 'S' equ.(\ref{gcond2}) becomes comparable to the 'NR' condition (\ref{pressureless}) at masses $\gtrsim (30-80)$ eV$/c^2$, quite at the limit of the phase-space bound (see equ. (\ref{massboundphasespace}) and text below). 
Nevertheless, we note that condition 'S' provides us with an \tx{a
posteriori} justification for the validity of the SI
regimes that we consider, which we will be able to appreciate in Section \ref{sec:4}. Furthermore, we see that ultralight particles $m \lll 1$ eV$/c^2$ demand a much stricter bound on $g$ in
 order to fulfil the 'NR' condition, than do particles of higher mass, which makes sense. In fact, we can see that the 'NR' bound (and 'V') are much stronger for ultralight particles than any of the GPP conditions. Therefore, we exclusively plot the ultralight boson mass regime, between $(10^{-22} - 10^{-6})$ eV/c$^2$, in the lower-right panel of Fig.\ref{fig1} and focus entirely on the 'NR' and 'V' bounds (for the other ones lie outside this plot range in $|g|$ and are automatically fulfilled once 'NR' and/or 'V' are fulfilled). However, we plot the bounds for two cases of halo core radii, $R_c = 1$ kpc and $R_c = 1$pc. The smaller the core, the smaller the remaining allowed parameter space for $|g|$, i.e. the curves shift correspondingly downwards. It so happens that the curves for 'V' ($1$ kpc) and 'NR' ($1$ pc) lie visually almost on top of each other which is a coincidence.

Before we move to the next section, let us make a final comment on the 'NR' condition. The argument concerning pressure versus energy density presumes that we can describe BEC-DM as a fluid, associated with a re-formulation of the GPP equations into fluid equations. In this sense, the 'NR' condition is of a different kind than the micro-conditions from the GP wave equation. In other words, it is assumed that the two pictures are equivalent. While not rigorously proven, there is plenty of evidence in the literature that this assumption is justified.

\section{Comparison to conditions from virial equilibrium}
\label{sec:3}

In this section, we put the results of Section \ref{sec:2} into perspective by comparing them to the conditions on BEC-DM parameters
implied by the assumption of virial equilibrium. In general,
once we demand that DM is able to form self-gravitating, bound galactic structures, we can
use dynamical arguments involving the equilibrium state of such
galactic halos. By the same token, however, we must remember that the constraints thereby derived on the DM parameters do depend on the assumption of equilibrium, which is approximately true at best. However, for certain halos or halo parts, such as their central cores, the equilibrium assumption may well be justified.  

We stress that the previous
 conditions (\ref{cond1}) and (\ref{cond2}), or equivalently (\ref{gcond1}) and (\ref{gcond2}), remain valid
 as long as the gravitational potential induced by the BEC-DM mass density does not vary much over the scales of the
 particle interaction range, which is the case here of the contact-SI. 
 On the other hand, equilibrium
 conditions of balancing forces on the scale of a given halo
 do impose constraints on the particle
 parameters, here $m$ and $g$. Especially the virial theorem has been widely applied to infer constraints on DM parameters, in general.
 
In the past, the equilibrium size of hydrostatic BEC-DM halo cores has been analysed in two limit regimes, where gravity is opposed predominantly either by quantum
pressure, or by the polytropic pressure due to the strongly
repulsive SI. Ref.\cite{RS12} includes an early comparison; these two regimes were studied there under the header ``BEC-CDM'' of type I and type II, respectively. FDM without SI, i.e. $g=0$, or with very weakly interacting bosons fall into type I. At the other end is BEC-DM with strongly repulsive SI (type II), more recently referred to as SFDM-TF in Refs.\cite{Taha21,Shapiro21}, or SIBEC-DM in Ref.\cite{Hartman21,Hartman22}. 
We discussed the conditions on pressure of BEC-DM to be nonrelativistic already in Section \ref{sec:2}. Still, BEC-DM pressure can be non-negligible on galactic halo scales, and this feature motivates BEC-DM as a potential remedy to resolve the small-scale crisis of CDM, as mentioned in Section \ref{intro}, because it provides a minimum scale, below which structure formation is suppressed, which can be much larger than for CDM. This statement, however, relies heavily on the parameters of the model, which are subject to observational constraints.
From theory, we can only generally state that quantum pressure stabilizes halo cores against gravitational collapse, while for SFDM-TF it is the polytropic pressure.
By the same token, either quantum or SI pressure will prevent formation of structure below a corresponding Jeans scale. For SFDM-TF, the SI-pressure determines that minimum scale \cite{Shapiro21}, as opposed to the de Broglie length for FDM. Again, observations constrain the size of these length scales.


Now, the smallest self-bound equilibrium object in BEC-DM serves also as a model for halo cores, whose radius and mass we denote again with $R_c$ and $M_c$. The general properties of this equilibrium object, especially its $M_c-R_c$-relationship and how it depends on BEC-DM particle parameters, have been known for a while, see e.g.\cite{Chavanis2011} (and references therein) which also includes some historic exposition. We present some of these fundamental results, before we derive the virial relationship that connects $m$ with $g$.

In general, the GPP system of equations does not admit closed-form equilibrium solutions, except for some limit cases concerning SI. However, variational estimates have been found and the corresponding analytic expressions often constitute good approximations of the true equilibria solutions. For example, a Gaussian for the spherical BEC-DM core density distribution
\beq \label{Gauss}
\rho_c(r) = \f{M_c}{(\pi R_c^2)^{3/2}}e^{-r^2/R_c^2}
\eeq
has been widely applied to the GPP system of equations, or the corresponding stationary energy functional, as an ansatz or test function, respectively, and it has been shown that it yields the same particle and halo parameter dependencies in the $M_c-R_c$-relationship than the exact solution, with numerical prefactors deviating only by numbers of order $\lesssim 1-10$, see e.g. \cite{2019PhRvD.100h3022C,2018PhRvD..97k6003G,Padilla20}. 
Using that Gaussian ansatz, the mass-radius relationship of a spherical equilibrium halo core is 
\beq \label{massradius} 
R_c = 3\sqrt{\f{\pi}{2}}\left(\f{\hbar}{m}\right)^2\f{1}{GM_c}
 \left(1 + \sqrt{1+\f{g}{3\pi^2}\left(\f{GM_c}{m}\right)^2\f{1}{G}\left(\f{m}{\hbar}\right)^4}\right),
\eeq
which is a re-written version of equation\footnote{Note that there is a typo in that equation; there should be no square in the last factor under the root.} 28 in \cite{Padilla20}.
Since the Gaussian has no compact support, the core radius is connected via the width of the Gaussian to the radius $R_{99}$, which includes $99\%$ of the core mass $M_c$, according to $R_{99} = 2.38167 R_c$. The relationship (\ref{massradius}) is valid for positive and negative\footnote{There exists a maximum mass $M_{c,max}$ in the attractive case $g < 0$, above which the system will undergo collapse. This has been known for laboratory BECs experimentally and theoretically, see e.g. \cite{JZ2000}, but has been also established in astrophysical contexts, see e.g. \cite{GHC15,2016JHEP...02..028E,2019PhRvD.100l3506C}. Also, in order for $R_c$ to be real, we require the condition $|g| \leq 3\pi^2Gm^2\left[\left(\f{1}{GM_c}\right)\left(\f{\hbar}{m}\right)^2\right]^2$.} SI-couplings $g$.
For FDM without SI, $g=0$, the mass-radius relationship of the so-called ``solitonic core'' (or ``soliton'') is
\beq \label{fdmrad}
R_{c,FDM} = 6\sqrt{\f{\pi}{2}}\left(\f{\hbar}{m}\right)^2\f{1}{GM_{c,FDM}},
\eeq
where the prefactor $6\sqrt{\pi/2} \approx 7.52$ differs not much from that of the numerically calculated soliton of $9.946$, see Ref.\cite{Membrado}. The core size goes inversely proportional to the core mass (where we added the subscript ``FDM''). Also, the smaller the boson mass $m$, the larger the core which, for ultralight bosons $m \sim 10^{-22}$ eV/$c^2$, is of order kpc, a favourable scale to possibly resolve the small-scale crisis. On the other hand, the core size and its dynamical impact in the halo is small for $m \gg 10^{-22}$ eV/$c^2$, and so the difference to CDM diminishes to a point where it is increasingly difficult to distinguish the models observationally.
In fact, the weakest lower-bound estimates for $m$ are found in FDM models without SI, as follows.
The smallest equilibrium size of
hydrostatic FDM halo cores is governed only by quantum pressure
and set by $\lb$ in (\ref{dB}), evaluated using the 
virial velocity of these cores. Since the smallest such objects also have the smallest
virial velocities, the most stringent bound results, if we require
$\lb \lesssim R_{c,FDM}$, where $R_{c,FDM}$ depends on the other parameters according to equ.(\ref{fdmrad}). Thus, a lower bound on $m$ is implied for fixed core mass $M_c$. Suppose we take $R_{c,FDM} = R_{vir}=1$
kpc and $M_c = M_{vir} = 10^8~M_{\odot}$ as proxy numbers for a fiducial dwarf-spheroidal-galactic halo that we postulate to host the smallest galaxies. This yields
 \beq \label{mlower}
 m \gtrsim (1-2) \times 10^{-22} ~\rm{eV}/c^2,
 \eeq
 using (\ref{fdmrad}) (or (\ref{mmin}) below).
This rough argument shows how the lower end of desired boson masses comes about: it is driven by the observational necessity for BEC-DM to account for the smallest galaxies that are known. Of course, the smallest galaxies may have yet to be found, let alone the smallest DM halos - by definition hard to accomplish and only possible indirectly for the latter. Therefore, bounds are provided by observations of ever smaller galaxies; favoured are now ultra-faint dwarf galaxies, which contain only hundreds to a few thousand stars. Recent works have used observations of such ultra-faint dwarfs in the Local Group in order to constrain the boson mass, and these yield lower bounds around $m \sim 10^{-21}$ eV$/c^2$ (see \cite{Nadler21,MN19}) which are $1-2$ orders of magnitude higher than the mass in (\ref{mlower}), explored in the early FDM literature. Such boson masses imply sub-kpc core size of $\lesssim 0.5$ kpc.

At the opposite end, we have the strongly repulsive TF regime. For this SFDM-TF, the second term under the square root in (\ref{massradius}) is much larger than one, so we can approximate 
\beq \label{tfrad}
R_{c,TF} \simeq \sqrt{6}\sqrt{\f{g}{4\pi G m^2}}
\eeq
(adding now a subscript ``TF'').
Different from the prefactor $\sqrt{6}\approx 2.45$ derived from the spherical Gaussian ansatz, a more accurate calculation would reveal\footnote{A homogeneous-sphere approximation would give the prefactor as $\sqrt{15/2}\approx 2.74$. Even if rotation of the core is included to account for typical spin parameters due to halo angular momentum, the prefactor remains close to $3$, see \cite{RS12} for details.} the prefactor $\pi$, appropriate for a spherical $(n=1)$-polytrope which has a compact support upon choosing the first zero of the density profile as the boundary radius of the sphere.  
For SFDM-TF, the core size does not depend on the core mass, and it is fixed once the ratio $g/m^2$ is fixed. The larger $g/m^2$, the larger the core, which can be of order kpc. But as for FDM, newer theoretical constraints demand radii smaller than kpc-size also for SFDM-TF, as we will discuss shortly. 

The above relationships are based on the assumption that the considered halo cores with radius $R_c$ and mass $M_c$ are in virial (even hydrostatic) equilibrium. As such, the corresponding equilibrium solutions are asymptotic states of the dynamical evolution for $t \to \infty$. Given that galactic halos are subject to mass infall from the ambient environment, including interactions and mergers with neighboring halos, we can use these cores as approximative models for the central parts of halos. The halo envelopes that enshroud those cores may be subject to perturbations and ongoing relaxation processes. Yet, a sufficiently isolated halo (core plus envelope) could still be in approximate virial equilibrium (if not in hydrostatic equilibrium), and this assumption is commonly made in order to estimate the dynamical mass of halos of galactic or galaxy-cluster size.
Whatever the detailed dynamical state of a given halo, we will focus on their central cores and assume they are in virial equilibrium, an assumption that is justified from simulations \cite{Taha21,Shapiro21,Hartman22}.
With this assumption on the cores, relationships between $m$ and $g$ are implied, that are independent of the dynamical halo state at large and more generally valid as a result, because there is no freedom left to pick different values for $m$ or $g$ in the halo envelopes. 

Before we derive the virial condition, let us introduce characteristic units for the particle
mass and SI coupling strength, defined in \cite{RS12}
and \cite{RS_Proc}, as follows
 \beq \label{mmin}
    m_H \equiv \f{\hbar}{R_c^2(\pi G \bar \rho)^{1/2}}
    = 1.066 \times 10^{-22}\left(\f{R_c}{1 ~\rm{kpc}}\right)^{-1/2}\left(\f{M_c}{10^{8}~ M_{\odot}}\right)^{-1/2}
  \rm{eV}/c^2,
    \eeq
     and
   \beq \label{gmin}
    g_{H} \equiv \f{\hbar^2}{2\bar \rho R_c^2}
    = 2.252 \times 10^{-62} \left(\f{R_c}{1~
   \rm{kpc}}\right)\left(\f{M_c}{10^{8}~M_{\odot}}\right)^{-1}
   \rm{eV}\rm{cm}^3.
     \eeq
We described the physical meaning and significance of these parameters
at length in \cite{RS12}. We only re-iterate here that $m_H$ is
the characteristic mass of a non-interacting particle whose de
Broglie length $\lb$, equ.(\ref{dBfidu}), is comparable to
the radius $R_c$ of a given BEC-DM halo core of mass $M_c$. It is thus the smallest particle mass
possible in order for quantum pressure to be solely responsible for
holding that core up against gravitational collapse; in essence this provides the lower-bound estimate on mass $m$ in the FDM regime, see again equ.(\ref{mlower}).
On the other hand, if there are density variations in the BEC-DM on the scale of the core radius $R_c$, then $g_H$ is the coupling strength for which the
quantum and SI pressure force terms are equal, for BEC-DM models with SI.

In the TF regime, it has been shown already in
\cite{RS12} that virial equilibrium of halo cores demand relationships of the form 
  \beq \label{virial}
  \f{g}{g_H} = const.\left(\f{m}{m_H}\right)^2,
  \eeq
  where the constant prefactor of order one depends again on the details of the core geometry (in \cite{RS12} a relation such as (\ref{tfrad}), but for an ($n=1$)-polytrope has been used). 
Also, an expression for the halo core radius has been derived in \cite{RS12} in the form of equation 46 there, which is $R_c = \f{\sqrt{3}\pi^{1/4}}{12} (g/g_H)^{1/2}\lb$. That means, $\lb$ can be arbitrarily small compared to
  $R_c$, if $g \gg g_H$. 
  Indeed, we can now appreciate why it was important to check the condition (\ref{gcond2}) -- aka (\ref{cond2})-,
  in order to guarantee that the scattering length is always much smaller than the de Broglie length, $|a_s| \ll \lb$, even in the ``extreme'' TF regime where $g \ggg g_H$ and $\lb \lll R_c$. 
  
  Using (\ref{virial}), $g \gg g_H$ implies $m \gg m_H$ in the TF regime. Notice, though, that the conditions 'D', 'S' and 'NR' described in the previous section are blind to virial equilibrium and to these inequalities.

\begin{figure*}
\centering
  \includegraphics{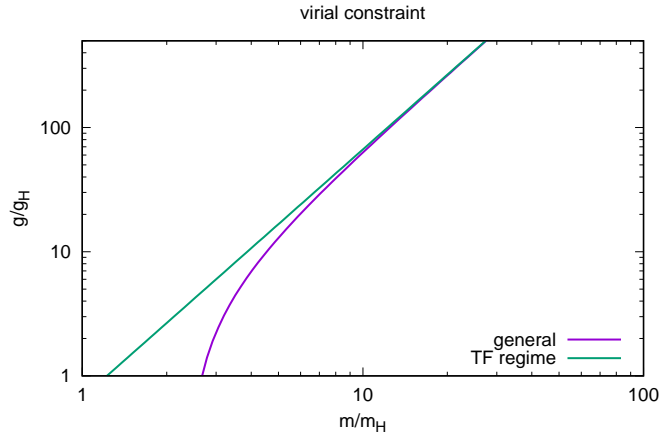}
 \caption{Virial relationships, equ.(\ref{gvirial}) (``general'') and equ.(\ref{tfvirial}) (``TF regime''). The curves lie on top of each other for $m/m_H \gg 1, g/g_H \gg 1$.}
 \label{fig2}
\end{figure*}

In light of the mass-radius relationship of the core in (\ref{massradius}), we can finally derive here the corresponding general virial condition. Inserting (\ref{mmin}) and (\ref{gmin}) into (\ref{massradius}) yields an implicit function of $m/m_H$, unlike the explicit function obtained in \cite{RS12}. However, in terms of $g/g_H$, we can derive an explicit equation which reads
\beq \label{gvirial}
\f{g}{g_H} = \f{54\pi}{32}\left(\f{m_H}{m}\right)^2 
\left(\left[\left(\f{m}{m_H}\right)^2 - \f{9}{4}\sqrt{\f{\pi}{2}}\right]^2\left(\f{4}{9}\right)^2 \f{2}{\pi} -1 \right).
\eeq
If $m/m_H \gg 1$, equ. (\ref{gvirial}) reduces to
\beq \label{tfvirial}
\f{g}{g_H} = \f{54}{81}\left(\f{m}{m_H}\right)^2,
\eeq
valid in the TF regime\footnote{We note that the prefactor differs slightly from the polytrope-based result in \cite{RS12}, because our expression here is based upon the Gaussian ansatz (\ref{Gauss}).}, see (\ref{virial}).
We plot both relationships, (\ref{gvirial}) and (\ref{tfvirial}),
in Fig.\ref{fig2} for comparison. The curves lie on top of each other in the validity range of the TF regime when $m/m_H \gg 1, g/g_H \gg 1$, as expected. In fact, the parameter space of SFDM-TF overlaps to such a great extent with the general parameter space of BEC-DM with repulsive SI, that we use relationship (\ref{tfvirial}), denoted as 'V', in Fig.\ref{fig1} altogether. The virial relation is a function $g=g(m)$, in contrast to the strong inequalities from GPP microphysics. It yields tighter constraints on $m$ and $g$ than the microphysical conditions, for $m \lesssim 100$ eV/$c^2$ which is just the range of interest given the phase-space condition in (\ref{massboundphasespace}). Thus, BEC-DM models, for which (halo) cores fulfil virial equilibrium, will automatically fulfil the micro-conditions from GPP.
   
Before we leave this section, we discuss some implications from recent literature, concerning the expected core size in SFDM-TF models. While the resolution of the CDM small-scale crisis demands core radii of order $\gtrsim 1$ kpc as shown in \cite{Taha21,Pils}, the recent studies of linear structure formation by \cite{Shapiro21,Hartman21,Foidl} find that the core radius is more tightly constrained than previously believed: not only are kpc-size cores disfavoured, it strongly appears that their radii should be below $\lesssim 100$ pc, if not even of order $\sim (1-10)$ pc. We refer the reader to these papers for details.
For comparison and in light of these findings, we thus add a constraint plot in the case of a very small halo core of $R_c = 1$ pc into Fig.\ref{fig2}, left-hand panel. There are no substantial differences to the previous cases shown in Fig.\ref{fig1}, i.e. the GP microphysical conditions do not limit our choice of core size. 
How small a core can we eventually choose? Ref.\cite{Padilla20} analyzed the question for which BEC-DM particle parameters halo cores could become unstable with respect to general-relativistic gravitational collapse, forming central galactic black holes as a result, and thereby possibly explaining the presence of supermassive black holes in the centers of galaxies at high and low redshift.
We can use their\footnote{We effectively use the results of \cite{Padilla20} in their Sec.V B, including their equation 94. Note that there is a typo in that equation: the factor in front should read $10^{25}$, instead of $10^{37}$. However, this typo does not affect the bounds and conclusions derived in \cite{Padilla20}.} condition of \tx{non-formation} of such a central black hole from a core for SFDM-TF models. In our notation, it reads as $g/(m^2c^4) \geq 1.23 \times 10^{-29}$ eV$^{-1}$ cm$^3$. Translating this into a bound on the core radius, we get that the halo core will not be subject to collapse as long as $R_c \geq 2.76 \times 10^{-3}$ pc $\approx 570$ AU.

\section{Scattering cross sections and collision regimes}
\label{sec:4}

While lower-bound estimates on particle mass $m$ of BEC-DM often invoke equilibrium/virial conditions applied to halos or their cores (see (\ref{mlower})), 
it has been argued that an \textit{upper} bound on $m$ arises by using the scattering cross section in the TF regime.  
For this purpose, it is useful to find an upper bound on the 2-particle scattering cross section \textit{per
particle mass}, $\sigma_s/m$ with $\sigma_s$ in (\ref{sigma}). 
Independent of the SI regime, it has been pointed out by \cite{SG} and \cite{RS12} that $\sigma_s/m$ may be re-expressed, using (\ref{coupling}) and
(\ref{sigma}), according to
 \beq \label{sigma2}
  \f{\sigma_s}{m} = \f{g^2 m}{2\pi \hbar^4},
  \eeq
 or in ``fiducial units'', using (\ref{mmin}) and (\ref{gmin}),
  \beq \label{sigma3}
   \f{\sigma_s}{m} =
    3.20 \times 10^{-95} \f{m}{m_H}\left(\f{g}{g_H}\right)^2 \left(\f{R_c}{\rm{kpc}}\right)^{3/2}\left(\f{M_c}{10^{8}
    M_{\odot}}\right)^{-5/2} \rm{cm}^2/\rm{g}.
    \eeq
   It has become customary in the literature on self-interacting DM to choose cgs-units for $\sigma_s/m$ (note: ``g'' stands for gram and shall not be confused with the coupling strength $g$), though we will see shortly that these units are not appropriate for BEC-DM. The canonical example of self-interacting DM, termed ``SIDM'', encompasses a family of different particle models. In a dynamical context, the most minimal difference of SIDM to CDM is the addition of a finite scattering cross section, modifying the usual N-body gravitational dynamics to include the physics of non-gravitational particle (elastic) collisions. While we are also concerned here with an elastic scattering cross section between BEC-DM bosons, the implications on $\sigma_s/m$ are different than in SIDM, as we show in this section. For this purpose, we will also need to make reference to some results pertaining to SIDM.
   
   First, we introduce the well-known Knudsen number ``$Kn$'' as a measure that distinguishes the fluid regime of strong SI (many particle collisions) versus the kinetic regime of few collisions (if any during the relevant time scale of the problem).
   Now, we use the inverse of the elastic scattering cross section $\sigma_s$ in
equ.(\ref{sigma}) to define the ``mean-free path'', i.e. the average distance travelled by a particle between two scattering events,
\beq \label{mfp}
\lmf \simeq 1/(n \sigma_s),
\eeq
where $n$ denotes again the particle number density. We stress that this relationship assumes that the system is sufficiently dilute, such that equations (\ref{cond1}) or (\ref{gcond1}) hold. So, the particles are supposed to interact only upon a scattering/collision event, when the separation between two particles is \textit{not} much larger than the interaction range 
$|a_s|$. This picture is justified for short-range SI potentials, such as the contact-SI upon which the GP formalism in equations (\ref{gp})-(\ref{coupling}) is based.
Now, the Knudsen number is the ratio of $\lmf$ over the characteristic size of the system, which could be the halo virial radius, or more appropriately in our context, the halo core radius. So, we write $Kn = \lmf/R_c$. The regime of $Kn \gg 1$ describes a system with few collisions (``Knudsen gas limit''), i.e. long mean-free path, while $Kn \ll 1$ indicates a system with short mean-free path.
Depending on the DM and halo parameters, either regime may be at work, in principle. In fact, for SIDM one encounters both regimes as shown in \cite{AS}, where a ``sweet spot'' of maximum (core-)flattening of the halo central density has been identified, corresponding to a certain Knudsen number, or equivalently to a ``$Q$-parameter'' introduced by the authors and which goes proportional to $R_{vir}/\lmf$.
They find that SIDM in the low-$Q$ regime, equivalent to a long mean-free path $Kn \gg 1$, behaves kinetic, and flattening of halo density cores becomes stronger as $Q$ increases. This regime can be studied by modified $N$-body dynamics, with typical values of $\sigma_s/m \simeq (0.1-10)$ cm$^2$/g. On the other hand, the high-$Q$ regime corresponds to short mean-free path $Kn \ll 1$, and central density flattening becomes weaker as $Q$ increases. This fully collisional regime of SIDM behaves like ordinary gas dynamics, with typical numbers\footnote{A complication arises in that SIDM model fits to observational data actually seem to favour different $\sigma_s/m$, depending on astronomical scale, where they should decrease as a function of relative velocities $v_\text{rel}$ of the DM particles. Taking $v_\text{rel}$ as a proxy for the virial velocity, fits of SIDM to galactic data on scales of $\sim (1 - 10)$ kpc with $v_\text{rel} \sim (50 - 200)$ km/s yield $\sigma_s/m \sim 1$ cm$^2$/g, while fits to data of galaxy clusters on scales of several Mpc and relative velocities of order $\sim 1000$ km/s generally yield smaller values of $\sigma_s/m \sim 0.1$ cm$^2$/g (see e.g. \cite{SIDM_scatt} for some discussion and the theoretical calculation of the scattering cross sections). In any case, the dynamical results in \cite{AS} extend upon the common lore that ``small'' $\sigma_s/m \lesssim 1$ cm$^2$/g are theoretically mandatory.} of $\sigma_s/m \approx (100-200)$ cm$^2$/g.    

Which regimes do we encounter for BEC-DM ?

Let us look at equ.(\ref{sigma3}) and consider first BEC-DM models with weak SI where $m/m_H \gtrsim 1$, $|g/g_H| \gtrsim 1$. We can readily see that in these cases $\sigma_s/m$ is a tiny number close to zero, for any reasonable choice of $R_c$ and $M_c$.
Such models behave de facto collisionless and by the same token the scattering cross section per particle mass is useless as a means to constrain the particle parameters. 
The smallness of $\sigma_s/m$ in this case comes with no
surprise as soon as we realize that BEC-DM with weak SI is equivalent to $Kn \gg 1$, i.e.
the Knudsen gas limit, where $N a_s^2 \to 0$ with $N$ the number of
bosons. 

\begin{figure*}
 \centering
 \includegraphics{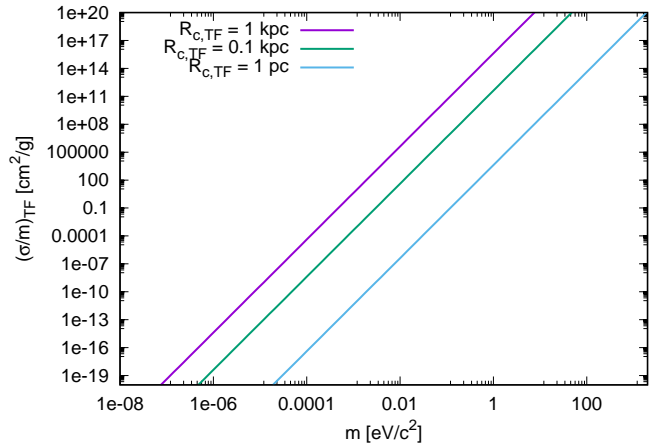}
 \caption{Specific scattering cross section in the TF regime as a function of boson mass $m$ for three different TF core radii $R_{c,TF}$.}
 \label{fig3}
\end{figure*}

At the other extreme, we have the strongly interacting TF regime of BEC-DM -- i.e. SFDM-TF-, which resembles the usual
Boltzmann gas limit with $N a_s^2 = constant$, and a significant boost of $\sigma_s/m$ as well as $Kn \ll 1$ is possible.
In these models, the characteristic size of interest is the radius of the TF core, $R_{c,TF}$ in (\ref{tfrad}) (remember that, in this regime, $R_{c,TF} \gg \lb \approx R_{c,FDM}$ for given $m$). 
Then, we can directly relate $\sigma_s/m$ to $R_{c,TF}$ by combining equations (\ref{coupling}), (\ref{sigma}) and (\ref{tfrad}), according to
\begin{equation} \label{cross}
\left(\frac{\sigma_s}{m}\right)_{\rm{TF}} = \frac{8 \pi
G^2}{36\hbar^4}R_{c,TF}^4 m^5,
\end{equation}
as first pointed out by \cite{SG} (but notice again the different prefactor that arises, if an $(n=1)$-polytrope is considered versus the Gaussian ansatz that we use here). This re-expression of the cross section is very helpful, given that neither the scattering length $a_s$, nor the coupling strength $g$ are direct observables, in contrast to the core radius $R_{c,TF}$ that can be constrained by galaxy data. Again, using ``fiducial units'', equ.(\ref{cross}) becomes
\beq \label{crossfid}
\left(\frac{\sigma_s}{m}\right)_{\rm{TF}} = 4.12 \times 10^{15} \left(\f{R_{c,TF}}{\rm{kpc}}\right)^{4}\left(\f{m}{\rm{eV}/c^2}\right)^{5} \rm{cm}^2/\rm{g}.
\eeq
In Fig.\ref{fig3}, we plot this quantity as a function of mass $m$ for three fixed values of the TF core radius $R_{c,TF}$. We notice that $(\sigma_s/m)_{\rm{TF}}$ can take on values that are orders of magnitude below or above typical SIDM numbers of $\approx (0.1 - 100)$ cm$^2$/g, due to the high power in $m$. Moreover, for bosons with $m \lesssim 10^{-4}$ eV/c$^2$, the scattering cross section per mass is still very small, even in the TF regime, hence for most of the parameter space of interest in $m$, BEC-DM is in the long mean-free path limit.
Finally, the larger the core radius, the higher $(\sigma_s/m)_{\rm{TF}}$ at given $m$. This effect is most likely of no observational consequence for bosons with $m \lesssim 10^{-4}$ eV/c$^2$, given the small numbers involved, as just pointed out. However, the predicted high numbers for $(\sigma_s/m)_{\rm{TF}}$ for bosons with $m > 10^{-4}$ eV/c$^2$ could prove useful in constraining such models in the future, comparing e.g. realistic halo merger simulations to observations
\footnote{    
Relations (\ref{cross}) and (\ref{crossfid}) implicitly assume that $m/m_H \gg 1$ -- in order to be in the TF regime in the first place-, in which case we noted already in \cite{RS12} (see equation 142 there), that (\ref{cross}) can be
rewritten, using the virial relation that connects $m$ with $g$. We can recover that 
formula here simply from (\ref{sigma3}) by inserting (\ref{tfvirial}) for $g/g_H$,
which yields
 \beq \label{sigmaTF}
  \left(\f{\sigma_s}{m}\right)_{\rm{TF}} = 
  5.65\times
  10^{-95}\left(\f{m}{m_H}\right)^5\left(\f{R_{c,TF}}{\rm{kpc}}\right)^{3/2}\left(\f{M_c}{10^{8}
  M_{\odot}}\right)^{-5/2} \rm{cm}^2/\rm{g}.
  \eeq
  In any case, the derivation of equ.(\ref{crossfid}) does not presume the virial condition $g=g(m)$, but the results are not very different if we had used (\ref{sigmaTF}) instead, in Fig.\ref{fig3}.}.
  
What shall we make with the result in Fig.\ref{fig3} ? Or phrased differently: What value or (upper) bound for $\sigma_s/m$ is expected from astronomical observations, such that we can exclude parameter space regions in that figure ?
In fact, there is no simple answer, because the inferred bounds from observations depend upon fitting the data to the underlying theoretical model of DM. The latter requires realistic simulations that are mostly still lacking for BEC-DM. Therefore, bounds from SIDM have been usually applied as a mere guess. For instance, it has been argued in \cite{SG}, and earlier in \cite{MUL2002}, that upper bounds on $\sigma_s/m$ for SIDM, based upon comparing that model to
observations, should apply to BEC-DM as well. This is a non-trivial assumption, though it may seem reasonable at first sight.   
Notably, the interpretation of the Bullet cluster observations as a
nearly collisionless merger of two cluster-sized halos has been
found to limit $\sigma_s/m$ for SIDM halos to $(\sigma_s/m)_{max} < 1.25$ cm$^2$/g, according to \cite{randall}, and therefore
a rough limit of $\sigma_s/m \lesssim 1$ cm$^2$/g has been invoked as a ``standard'' bench mark. 
In fact, by pursuing this SIDM analogy argument and by applying that bench mark in \cite{SG, MUL2002} (an argument which was re-iterated in \cite{RS_Proc,RS14}), an \textit{upper}
bound on $m$ follows readily upon a fiducial choice of halo parameters: e.g. using here formula (\ref{sigmaTF}) with $R_{c,TF} = 1$ kpc and $M_c = 10^8~ M_{\odot}$, we have $m \lesssim 7.5 \times 10^{-4}$ eV/$c^2$,
 in good accordance with the result in \cite{SG}, see their equation 6.
 Coincidentally, this mass scale is ``close'' to the QCD axion. However, if we choose a much smaller core radius of, say, $R_{c,TF} = 1$ pc, using the same central core mass $M_c$, we would get a higher upper-bound estimate of $m \lesssim 0.2$ eV/c$^2$. So, there is nothing fundamental about this ``upper bound''.
By comparing the SIDM benchmark with the results of Fig.\ref{fig3}, or with equ.(\ref{sigma3}) more generally, we can see that $\sigma_s/m$ is many, many orders of magnitude smaller
than $1$ cm$^2$/g, for any halo parameters, for ultralight bosons for which $m/m_H \gtrsim 1$, but \tx{also} for a large part of the TF regime where $m/m_H \gg 1$. It is only in the 
``extreme'' TF regime with $m/m_H \ggg 1$ that the cross section per mass is very high\footnote{As an illustration of some numbers, consider a laboratory atomic BEC gas. Take e.g. typical values from \cite{BECpaper} for $^{87}$Rb atoms. They have a mass of $m = 87~u = 1.44\times 10^{-25}~\rm{kg} = 8.1\times 10^{10}~\rm{eV}/c^2$, and a scattering length of $a_s \approx 98~a_0$ with the Bohr radius $a_0 = 5.3\times 10^{-11}~\rm{m}$, i.e. $a_s \simeq 5.2\times 10^{-7}~\rm{cm}$. Therefore, $\sigma_s = 8\pi a_s^2 \simeq 6.8\times 10^{-12}$ cm$^2$, while $\sigma_s/m \simeq 4.7\times 10^{10}$ cm$^2$/g.}, see Fig.\ref{fig3}: e.g. for $m = 0.1$ eV/c$^2$ and $R_{c,TF} = 1$ kpc, we have $\sigma_s/m = 4.1\times 10^{10}$ cm$^2$/g. 
 
There is another point on which we shall comment. Ref.\cite{SG} constructed a halo model, where a TF core of condensed BEC-DM described by an ($n=1$)-polytrope, is enshrouded by an envelope as an isothermal sphere made of thermal \tx{non-condensed} bosons. (The isothermal sphere provides rotation curves with constant velocity at large galactocentric radii.) By calculating the exact equation of state, that interpolates between core and envelope, the authors find that the transition region exemplifies a discontinuous feature that translates into a marked feature in the halo density profile. This feature is subject to observational constraints (via galactic rotation curves), concluding that, in order to avoid these constraints, the boson mass must fulfil contradicting lower and upper bounds, whereby the upper bound comes from the SIDM argument of forcing $\sigma_s/m \lesssim 1$ cm$^2$/g for the condensed core.
As we have seen, there is no fundamental reason to apply this bound to BEC-DM. However, the adopted requirement of the non-condensed bosons in the envelope to be in local thermal equilibrium demands that these bosons (as a thermal gas) be in the small mean-free path regime, i.e. $\lmf \ll R$, with $R$ a characteristic radius, equivalent to
$\sigma_s/m \gg 1/\rho R$, or 
 \beq \label{thermo}
 \f{\sigma_s}{m} \gg 1.82
 \left(\f{\bar \rho}{\rm{GeV}/(c^2\rm{cm}^3)}\right)^{-1}\left(\f{R}{100~\rm{kpc}}\right)^{-1}
 \rm{cm}^2/g,
 \eeq
 where $\sigma_s$ stands merely for an elastic cross section, not related to equ.(\ref{sigma}), and $\bar \rho$ denotes a mean mass density of thermal bosons in the envelope. 
  The number in (\ref{thermo}) is $\gg 70$ cm$^2$/g for the Milky Way at the solar circle, assuming that the sun's orbit is located within such an envelope, thus a much higher number arises than the enforced bound on this ratio in the core, which questions the significance of that core bound. 
  Indeed, the model construction in Ref.\cite{SG} seems not necessary, because i) later performed halo formation simulations of fully condensed BEC-DM have revealed that a core-envelope structure arises in halos in any case, with an envelope close in behaviour to ``isothermality'' (even without a component of thermal bosons), and the transitional density region between core and envelope need not be discontinuous at all, and ii) once we consider a halo of condensed BEC-DM, we have seen that the largest values for $\sigma_s/m$ will be found in the TF core, in the ``extreme'' TF regime where $m > 10^{-4}$ eV/c$^2$. As such, any constraints on models with high $\sigma_s/m$ must be placed on these cores by some other means, unrelated to a marked density feature in the core-envelope profiles of individual galactic halos. In addition, if TF cores are constrained to be small, of order $\sim (1-100)$ pc according to recent literature, the ratio $\sigma_s/m$ becomes small, too, making observational constraints yet again harder to place.   
  Concerning the work premise of considering halos of fully condensed BEC-DM (without thermal component), we elaborated in Section \ref{intro}, that the bosons of BEC-DM are assumed to have undergone a BEC phase transition much earlier than the formation time of halos, in order that we can apply the GPP formalism to halos of that BEC-DM in the first place. So, in this sense the premise is self-consistent.
  
Let us finally make a last comment to add another illustration of why using bounds on scattering cross section from SIDM to BEC-DM is misleading. Remember that 
 $\sigma_s/m \approx 1~ \rm{cm}^2/\rm{g} = 1.8~\rm{barn}/(\rm{GeV}/c^2$), or $\sigma_s = 1.8 \times 10^{-24}~\rm{cm}^2 \left(\f{m}{\rm{GeV}/c^2}\right)$, i.e. for typical SIDM particle masses of $\approx$ GeV/c$^2$, the scattering cross sections are similar to nucleons, which explains the choice of units.        
Now, let us use equation (\ref{crossfid}) - equivalently (\ref{sigmaTF}) - and write it in terms of $(\sigma_s)_{TF}$, yielding 
\begin{equation}
 \sigma_{s,TF} = 7.3 \times 10^{-18}~\rm{cm}^2\left(\f{m}{\rm{eV}/c^2}\right)^6\left(\f{R_{c,TF}}{\rm{kpc}}\right)^4.
\end{equation}
If we would like to have a typical SIDM (nuclear) cross section of $\sigma_{s,TF} \approx 10^{-24}$ cm$^2$, we would need a BEC-DM boson mass of $m \approx 0.1$ eV/c$^2$ if $R_{c,TF} = 1$ kpc, or $m \approx 10$ eV/c$^2$ if $R_{c,TF} = 1$ pc.  
On the other hand, fixing the same choice of core radii and considering ultralight boson masses, $m \approx 10^{-21}$ eV/c$^2$, implies a tiny number for $\sigma_{s,TF}$ close to zero, i.e. negligible.
Concerning the ratio $\sigma_s/m$, however, we have seen that it is always very small, if $m \lll 1$ eV/c$^2$, due to the dependence on $m^5$ in (\ref{crossfid}) which drives it to tiny numbers. Only the ``extreme'' TF regime, where $m > 10^{-4}$ eV/c$^2$, provides high $\sigma_s/m$.

Finally, we show that our conclusions in this paper do not change, if we include the dependence of the scattering cross
section on ``temperature'', which effectively here is equivalent to virial velocity, or another characteristic velocity of that order, such as $\vc$. 
Taking into account the effects of finite temperature ($k \not= 0$) in the
calculation of the 2-particle interactions means to go beyond
the first Born approximation in (\ref{coupling}) and (\ref{sigma}),
which, again, is a well-established result in the literature on
s-wave scattering at low energies. Including the first order term in
kinetic energy $E$ (or $k^2$, respectively) in the expansion of the
scattering amplitude for indistinguishable bosons, results in the following
modified equation for the scattering cross section
 \beq \label{sigmaT}
\sigma(k) = \f{8\pi}{k^2 + (-1/a_s + r_e k^2/2)^2},
 \eeq
  see e.g. \cite{joachain} and references therein. This formula is valid so
  long as the 2-particle interaction potential $W(\mb{r})$ under the integral in (\ref{scattlength2}) is less singular than $1/r^2$ around the
  origin, and vanishes faster than $1/r^3$ at infinity\footnote{These conditions guarantee the analyticity
  of the scattering amplitude as a function of $k^2$ (or $E$), and, hence, its expansion into a power
  series in $k^2$.}. The 'effective range' $r_e$, which appears in (\ref{sigmaT}), is defined in terms
of the free wavefunction and the scattering wavefunction at large
radii, respectively. In the zero-temperature limit,
 $k\to 0$, formula (\ref{sigmaT}) goes over into (\ref{sigma}). Using now
 $k^{-1}=\hbar/(mv) = \lb(\vc)/(2\pi)$ (by setting $v=\vc$), we rewrite equ.(\ref{sigmaT}) as
  \bdi
  \f{\sigma(\vc)}{m} = \f{8\pi a_s^2}{m}
  \left[1+(2\pi)^2\left(\f{a_s}{\lb(\vc)}\right)^2\left(1-\f{r_e}{a_s}\right) + \f{(2\pi)^4}{4}\left(\f{a_s}{\lb(\vc)}\right)^2\left(\f{r_e}{\lb(\vc)}\right)^2\right]^{-1},
   \edi
which reduces to equ.(\ref{sigma}) in the limit of small quantum parameter, see (\ref{cond2}) or (\ref{gcond2}). The scattering length $|a_s|$ is now no longer related to $g$ in the form of equ.(\ref{coupling}), and instead has become 
the first, albeit leading order term of the coupling strength; see also \cite{LL}. Also, we have $|a_s| \gtrsim r_e$ in general, as long as the SI is sufficiently short-range, which we assume, hence we can
estimate that
 \beq \label{sigmafinal}
  \f{\sigma(\vc)}{m} >  \f{\sigma_{\star}(\vc)}{m} \equiv 
  \f{8\pi a_s^2}{m}
  \left[1+(2\pi)^2\left(\f{a_s}{\lb(\vc)}\right)^2 +
  \f{(2\pi)^4}{4}\left(\f{a_s}{\lb(\vc)}\right)^4\right]^{-1},
  \eeq
using (\ref{sigma}) and (\ref{dBfidu}). The velocity-dependent cross section
$\sigma(\vc)/m$ decreases with increasing velocity, however,
the additional terms due to $\vc$ are negligible compared
to the leading-order term $\sigma_s/m = 8\pi a_s^2/m$ in (\ref{sigmafinal}), given the regime of small quantum parameter, $|a_s|/\lb(\vc) \ll 1$, that we presumed in Section \ref{sec:2}, and the fact that (\ref{cond2}) and the ensuing relationships are valid also in the extreme TF regime, as we have shown above. 


\section{Conclusions and Discussion}
\label{sec:5}

Bose-Einstein-condensed dark matter (BEC-DM), also called scalar field dark matter (SFDM) has become a popular alternative to the standard, collisionless cold dark matter (CDM) model. If the boson masses of BEC-DM are ultralight, $m \gtrsim 10^{-22}$ eV$/c^2$, the dynamics between BEC-DM and CDM differs on kpc scale, providing new signature effects and a possible resolution of the small-scale structure problems of CDM. However, in recent years, a collection of new constraints in the literature have been determined on ``fuzzy dark matter (FDM)'' models of BEC-DM, where the mass $m$ is the only free parameter. These bounds concern the structural length scales equ.(\ref{fdmrad}), pointing to the requirement of $m > 10^{-21}$ eV$/c^2$ which makes the distinction to CDM smaller than previously thought. On the other hand, BEC-DM with a self-interaction (SI) introduces a further parameter, the SI-coupling strength $g$, implying different constraints. For models in the Thomas-Fermi (TF) regime of strongly repulsive SI, also called SFDM-TF or SIBEC-DM, the characteristic structural length scale is again described by a single parameter combination, $\sim g/m^2$ see (\ref{tfrad}), which is subject to constraints. Regardless of regime, these length scales are also related to the typical core size of BEC-DM halos.

In this paper, we derived the conditions on the BEC-DM particle parameters which result from the fundamental microphysical assumptions, which underlie the equation of motion of BEC-DM used in the literature, the Gross-Pitaevskii (GP) equation coupled to the Poisson equation; the GPP framework. To our knowledge, it has never been checked before what these assumptions entail for the DM parameters. 
More precisely, we have derived upper bounds on $g$ which follow from the requirements that the system is sufficiently dilute (smallness of ``gas parameter'' (\ref{cond1}), aka condition 'D' of (\ref{gcond1})), and that only s-wave scattering is important (smallness of ``quantum parameter'' (\ref{cond2}), aka condition 'S' of (\ref{gcond2})). BEC-DM models without SI, $g=0$, or very weak SI are also covered thereby. Also, a bound on $g$ follows from the condition that BEC-DM pressure be negligible, compared to its energy density (i.e. behaving like nonrelativistic ``cold dust'', condition 'NR' of (\ref{pressureless}) on cosmological scales). We expressed our conditions on $g$ as a dependence on boson mass $m$, as well as on (mean) density and (circular) velocity of the system, which are considered free parameters. Here, we probed the conditions using various halo core parameters. The 'NR' condition provides the best constraint for most of the parameter space in $m$, although the condition 'D' becomes more important for $m \gtrsim (0.1-10)$ eV/c$^2$. Condition 'S' has barely any constraining power, but by the same token it implies (the \tx{a posteriori} justification) that s-wave scattering is the only relevant scattering process in BEC-DM for boson parameters of interest.   

Apart from the microphysical conditions, we also derived a relationship that connects $m$ and $g$ under the premise that virial equilibrium can be assumed for BEC-DM halo cores. We thereby extended a previous result of \cite{RS12} that was limited to the TF regime.
It turns out that this virial constraint is the most stringent condition on the BEC-DM parameters, for all $m \lesssim 100$ eV/$c^2$. So, this is an example where dynamical constraints are stronger than those from microphysics, \textit{if} the assumption of virial equilibrium is correct. 

We also revisited and derived the implications for the elastic scattering cross section per particle mass, $\sigma_s/m$, of BEC-DM, with and without the inclusion of velocity-dependence. In either case, $\sigma_s/m$ is a very small number, close to zero, for BEC-DM models with weak SI, but also for many models of interest in the TF regime. Only in the extreme TF regime where $m \gtrsim 10^{-4}$ eV/$c^2$, can $\sigma_s/m$ attain high values, also dependent on the assumed halo core size. 
In the process, we could show that the mere adoption of the ``self-interacting dark matter (SIDM)''-inspired bound of $\sigma_s/m \lesssim 1$ cm$^2$/g on BEC-DM models is not correct. In fact, depending on boson parameters, $\sigma_s/m$ can take on values that are many orders of magnitude below or above the typical numbers of SIDM, and a proper comparison of theoretically allowed values to observational galaxy data would require realistic simulations of BEC-DM, e.g. Bullet cluster-type halo collision simulations in the GPP framework.

We also put our results into context to previous literature, as follows.
By assuming the virial constraint, we have implicitly made use of the fact that BEC-DM halos generically form a core, surrounded by an envelope. The cores constitute basically the equilibrium states of BEC-DM (i.e. close to virialization), while the envelopes have been shown to be very dynamical, see e.g. \cite{SCB14,Schive14,SNE16,Mocz17} for FDM. One-dimensional halo formation simulations of SFDM-TF have been performed in \cite{Taha21}, where it has been established that a core-envelope structure also arises; in this case a TF core -- very close to an $(n=1)$-polytropic core- was found, enshrouded by an envelope which is CDM-like. The basic results were subsequently confirmed by 3D simulations in \cite{Hartman22}, who also find a dynamical envelope structure.

The halo envelopes found in both FDM and SFDM-TF simulations are similar to isothermal spheres.
Previously, it has been shown in \cite{SG} that, \textit{if} a TF core of condensed BEC-DM is enshrouded by an isothermal envelope of \textit{non-condensed} bosons in local thermodynamical equilibrium, the transition between the different equation of states between core and envelope would result in a very sharp feature in the halo density profiles. Since such a sharp feature would be in conflict with observed galactic velocity profiles, it was argued by \cite{SG} that either such a core-envelope model would be ruled out, or that it would require a boson mass in conflict with other bounds\footnote{In fact, any DM species whose equation of state changes sharply enough between inner halo region (``core'') and outer region (``envelope'') to produce observational features in the density and velocity profiles, could be constrained that way. Again, the statement in \cite{SG} applies only to bosons in thermal equilibrium; see also the discussion in \cite{2019PhRvD.100h3022C}, particularly footnote 20.}. 
However, contrary to the findings in \cite{SG}, the analysis of SFDM-TF in \cite{Taha21,Shapiro21,Hartman22} has revealed that a fatal sharp feature between core and envelope need not arise. The important difference is that there is only condensed BEC-DM throughout the halos in \cite{Taha21,Shapiro21,Hartman22} (core plus envelope), and the apparent ``isothermality'' of the envelope there results from a purely effective description of the complicated wave dynamics of BEC-DM. A proper coarse-graining of the quantum pressure of BEC-DM gives rise to a ``velocity dispersion pressure'', similarly to CDM.

Finally, recent works by Refs.\cite{Shapiro21,Hartman21,Foidl} have found that theoretical constraints from linear structure formation highly suggest that halo cores in SFDM-TF should be smaller than kpc size, even preferring core radii with $\lesssim (1 - 100)$ pc.
Therefore, we have also probed very small halo cores in this paper, finding that the GPP microphysical conditions are no obstacle to that choice; we are safe to pick such small radii. Like in the case of large cores, the virial condition provides the most powerful constraint on $g$ here as well.
From an observational perspective, small halo cores are very hard to infer and may be indistinguishable from core-free CDM models altogether. However, we stress that the question of the smallest-allowed core size in BEC-DM is not yet settled, given the computational difficulties of cosmological halo formation simulations. The latter are the subject of ongoing research in the community.

\section*{Conflict of Interest Statement}
 
The author declares that the research was conducted in the absence of any commercial or financial relationships that could be construed as a potential conflict of interest.



\section*{Funding}
This work was supported by the Austrian Science Fund FWF through an Elise Richter fellowship to the author, Projektnummer V656-N28.

\bibliographystyle{frontiersinHLTH&FPHY} 
\bibliography{Scattering_REV_ar}{}


\end{document}